\documentclass[aps,prl,amssymb,twocolumn,showpacs]{revtex4}

\usepackage{graphicx}

\newcommand{\bea}{\begin{eqnarray}}

\newcommand{\eea}{\end{eqnarray}}

\newcommand{\beq}{\begin{equation}}

\newcommand{\eeq}{\end{equation}}

\newlength{\textwidthm}

\begin{document}

\title{Creating a supersolid in one-dimensional Bose mixtures}

\author{L.~Mathey$^1$, Ippei Danshita$^2$, Charles W. Clark$^1$}

\affiliation{$^1$Joint Quantum Institute, National Institute of Standards and Technology,
 Gaithersburg, MD 20899\\
$^2$Department of Physics, Faculty of Science, 
Tokyo University of Science, Shinjuku-ku, Tokyo 162-8601, Japan}

\date{\today}

\begin{abstract}
We identify a one-dimensional supersolid phase in a binary mixture of near-hardcore bosons with weak, local interspecies repulsion. We find realistic conditions under which such a phase,
 defined here as the coexistence of quasi-superfluidity and quasi-charge density wave order, can be produced and 
 observed in finite ultra-cold atom systems in a harmonic trap. Our analysis is based on Luttinger liquid theory supported with numerical calculations using the time-evolving block decimation method.  
Clear experimental signatures of these two orders can be found, respectively, in time-of-flight interference patterns, and the structure factor $S(k)$
 derived from 
density correlations. 
%
\end{abstract}


\maketitle
The supersolid (SS) phase,
 defined as a many-body state that simultaneously shows superfluid (SF) 
 and charge density wave (CDW), i.e. crystalline, order,
  has been an intriguing notion since its first
 proposal \cite{earlysupersolid}, due to its
 seemingly paradoxical nature.  
  Numerous studies of
 SS phases 
 \cite{supersolid, commix} have recently been reported,
  motivated by 
   fundamental theoretical interest in a system that exhibits
    competing orders, and by
     recent experimental reports of observations of supersolidity  
  of $^4$He in vycor glass \cite{kimchan, kimchantheory}.
%
%
%
  This $^4$He system 
 exemplifies the complexity of studying 
 strongly correlated systems in a solid-state context: 
  it combines strong disorder due to the porous medium, 
 strong interactions between atoms and of atoms with surfaces.
   Under such circumstances 
  it is difficult to 
  demonstrate the existence of 
  a
 SS phase, which
  involves a subtle competition of fluctuations.
  
  In this paper we show that supersolidity can be studied with clarity in another 
   physical system: ultra-cold atoms in optical lattices.  
 Since the demonstration of the 
 SF-Mott insulator transition in 3D
 \cite{greiner},   
 the technology of cooling and trapping atoms has 
  supported studies of numerous quantum many-body phenomena, 
  such as BEC-BCS crossover \cite{BECBCS}, noise correlations \cite{noise},
  the Berezinsky-Kosterlitz-Thouless 
  transition \cite{zoran},
  the Tonks-Girardeau gas \cite{Tonks, kinoshita},
 transport and collisional properties of one-dimensional (1D)
  gases \cite{1Dgas}, 
 and the Mott
transition in 1D \cite{stoeferle} and 
 2D \cite{Ian}.
 Appealing features of 
 this technology, from the
 perspective of many-body theory, are that it creates well defined
 and tunable 
 systems, and that the set of measurable
 quantities differs from those in solid-state systems.
  Thus, ultra-cold atom systems can give interesting
 and unusual insights into many-body states.

The objective of this paper is to propose a realistic setup
 of how to create and detect a supersolid with current technology.
%
 Specifically, a binary mixture of near-hardcore bosons with weakly repulsive inter-species
 contact 
   interactions  in a 1D potential displays both 
 CDW 
      and 
 SF quasi-long range order (QLRO). 
  Such mixtures have an inherent tendency to undergo phase separation, which can be
   avoided if the inter-species interactions are sufficiently weak.


 We study the 
 SS phase
 with analytical and numerical techniques, and
  present a concrete proposal for its realization in current experimental systems.  
 First we use  
 a Luttinger 
 liquid (LL) approach
 to derive 
 the phase diagram of the
 homogeneous, infinite system, 
 with 
 a renormalization group (RG) calculation.
 We then 
 address the question of realizing such a phase under realistic
 conditions, 
 for a finite system of $\sim 10^2$ lattice sites 
  in 
  a harmonic trap. Using 
 a number-conserving 
 time-evolving
 block decimation (TEBD) method \cite{vidal} we numerically
 determine, with a well-controlled error, the ground state of 
 the system 
 from which we extract various correlation functions. 
  We first identify the 
 SS phase, through signatures
 in the pair and anti-pair correlations.
   Other correlations
 contain information that is accessible to direct experimental
    observation, and we discuss possible experimental signatures
     of the 
 SS phase,  
  i.e. the coexistence of SF and CDW order; 
   the SF order is manifest in the single-particle correlation function,  which can
    be determined from time-of-flight (TOF) interference patterns;
     the CDW order is seen in density-density correlations, which is reflected
      in a measurable structure factor.
 
We consider a mixture of two species of bosonic atoms with short-range interparticle interactions, confined in a 1D optical lattice and an additional harmonic potential. 
Contemporary experimental realizations of such systems are usually well
approximated  by a Hubbard model:
\bea\label{Hubbard}
H & = & - t \sum_{<ij>, a} b^\dagger_{a, i} b_{a, j}
 + \frac{U}{2} \sum_{i,a} n_{a,i} (n_{a,i} -1)\nonumber\\
& & + U_{12} \sum_{i} n_{1,i} n_{2,i} + \sum_j \Omega j^2 
 (n_{1,j} + n_{2,j}).
\eea
%
%
%
 Here $t$ is the hopping energy;
$b_{a,i}$ is a boson field operator, with $a=1,2$ a species index and $i$ a lattice site index;   
 $U$ ($U_{12}$) is
 the intra-  (inter-)species on-site interaction energy;
 $n_{a,i} = b_{a,i}^\dagger b_{a,i}$; and
 $\Omega$ represents the strength of the harmonic trap, which is centered
on the site $j=0$.

%
%
We now derive the phase diagram of this 
 system 
 from LL theory.
 We consider two 1D bosonic SFs, with densities equal
 to each other, but incommensurate to the optical lattice. 
 In particular we exclude half- and unit-filling, which would destroy the
  SF order, by choosing the density and the global
 trap in such a way that even at the trap center the density of each
 species stays below $0.5$.
  The essential function of the optical lattice 
  is to provide a 
   sufficiently large 
  ratio of $U/t$, 
 as 
  in [\onlinecite{Tonks}]. 
 We emphasize that the supersolid phase also exists
  in the absence of a lattice\cite{Schulz93};
  however, since present quasi-homogeneous 
  ultracold atomic systems are generically closer to weak coupling, the presence of a lattice is advantageous to experimental realization of supersolidity.
%
%
%
%
%
%
%
%
%
%
\begin{figure}
\includegraphics[width=7.8cm]{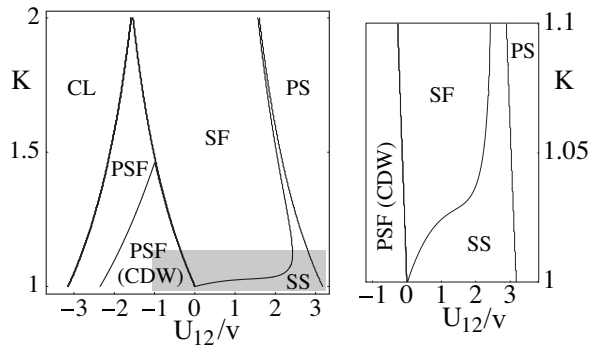}
\caption{\label{PD}
 Phase diagram of a Bose mixture, from LL theory,
 as a function of the 
 inter-species
interaction
 $U_{12}$ (in units of 
  $v$, see text), 
 and the LL parameter $K$ of the uncoupled
 system  
 (Eq. (\ref{K})); 
  on the left 
  the full diagram, on the right (corresponding to the
 shaded area) the near-hardcore, repulsive
 regime with the supersolid (SS) phase. 
  For attractive interactions,
 in the paired regime, 
  paired SF (PSF) is dominant,
 with 
 CDW QLRO being subdominant in parts of it.
  Outside of that regime, 
 SF is the dominant quasi-order, but
 for the repulsive, near-hardcore regime we have CDW QLRO as well, 
 which constitutes a supersolid phase. 
 For large repulsive interactions the system phase separates (PS),
 for large attractive ones
 it collapses (CL).
}
\end{figure}

The basic concept of LL theory is to express the bosonic operators $b_{a,i}$
  through a  
 bosonization identity, 
 such as Haldane's construction \cite{Haldane, Cazalilla}:
\bea
b_{1,2}(x) & = & [n + \Pi_{1,2}(x)]^{1/2} \sum_{m} e^{2 m i\Theta_{1,2}(x)} 
 e^{i\Phi_{1,2}(x)},
\eea
 where we switched to a continuum model, $b_{a,i}\rightarrow b_a(x)$. 
$n$ is the average density of the two species,
 $\Pi_{1,2}(x)$ are the low-k parts (i.e. $k\ll 1/n$) 
 of the density fluctuations; 
 the fields  
 $\Theta_{1,2}(x)$ are given by 
 $\Theta_{1,2}(x) 
= \pi n x + \theta_{1,2}(x)$, 
with $\theta_{1,2}(x)=\pi \int^x dy \Pi_{1,2}(y)$.
 $\Phi_{1,2}$ 
 are the phase fields, 
 the conjugate fields of the density fluctuations $\Pi_{1,2}(x)$.
%
%
%
%
%

In terms of these fields,  
  the action of the two coupled bosonic SFs is given by 
 \cite{giamarchi_book, Cazalilla}:
%
%
%
%
%
%
%
%
%
%
\bea\label{Shalf}
S & = &  \int d^2r \Big[\sum_{j=1,2} \frac{1}{2\pi K} \Big((\partial_{\tau} \theta_{j})^2 +   (\partial_x \theta_{j})^2\Big)\nonumber\\
 & & + \frac{U_{12}}{\pi^2} \nabla \theta_1 \nabla \theta_2
 +   \frac{2 g_{\sigma}}{(2 \pi\alpha)^2}
\cos(2\theta_1-2\theta_2) \Big]
\eea
%
%
%
%
%
%
%
%
%
%
%
%
%
%
%
%
%
The two SFs are 
characterized by a LL parameter $K$ and a velocity $v$, which 
 is contained in 
  ${\bf r}=(v \tau, x)$.  
 The LL parameter 
 $K$ is a measure of the intra-species interaction; 
 in the near-hardcore
 regime we use
 \cite{CazalillaTonks}:
\bea\label{K}
K & \approx & 1 + \frac{8 t}{U} \frac{\sin \pi n}{\pi}
\eea
Similarly, $v$ can be related to the
 parameters of the underlying Hubbard model by $v \approx v_F(1-8 t n \cos \pi n/U)$,
 where $v_F$ is the `Fermi velocity' 
  of an identical system of fermions, 
 $v_F=2 t \sin \pi n$.
%
%
%
%
%
%
%
%
%
%
%
%
%
%
%
%
 The density-density interaction between the two SFs
 creates both the term containing $\nabla\theta_1\nabla\theta_2$, 
%
%
%
%
%
%
%
%
%
%
%
%
%
%
%
%
as well as the backscattering term \cite{giamarchi_book, commix},
 containing $\cos(2\theta_1-2\theta_2)$,
  which describes short-range interspecies repulsion. 
 %
%
%
%
%
%
%
\begin{figure}
\includegraphics[width=5.2cm]{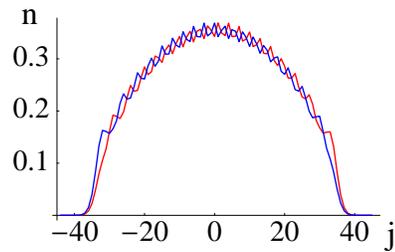}
\caption{\label{density}
 The single-atom density of species 1 (red)  and 2 (blue) 
 of a system of 90 sites, with 17 atoms
 of each type, with $t/U=0.005$, $U_{12}/U=0.04$,  
 and a trap parameter $\Omega=10^{-5}U$.
}
\end{figure}
%
%
%
%
%
%
 We change  
variables to the symmetric and
antisymmetric combinations
%
$\phi_{s/a} =  \frac{1}{\sqrt{2}} (\phi_1 \pm \phi_2)$ 
 and 
$\theta_{s/a} = \frac{1}{\sqrt{2}} (\theta_1 \pm \theta_2)$, 
%
%
 and diagonalize the quadratic part of the action which gives the 
 following parameters for the two sectors:
\bea
K_{s/a} & = & (1/K^2 \pm U_{12}/v\pi K)^{-1/2} 
\eea
 which to lowest order gives $K_{s/a}\approx K\mp U_{12}K^2/2\pi v$.
 The effective velocities are 
 $v_{s/a}=\sqrt{v^2\pm U_{12}K v/\pi}$.
 Phase separation (collapse) is reached when $v_{a (s)}$ becomes
 imaginary.  
The anti-symmetric sector contains the nonlinear 
 backscattering term. 
 To study its effect, we use an RG approach; 
%
%
 the flow equations for which are given by\cite{giamarchi_book}:
\bea\label{RG}
\frac{d g_\sigma}{d l} & = &(2-2K_a) g_\sigma; \,\,\, \frac{d K_a}{d l}  =  
- \frac{g_{\sigma}^2}{2\pi^2}  K_a^3 
\eea
%
%
%
%
This set of flow equations has two qualitatively different fixed
 points: Either $g_\sigma$ diverges, driving
 a pairing transition, which in turn renormalizes $K_a$ to
 zero, or $g_\sigma$ is renormalized to zero.
 In the latter case the Gaussian fixed point is restored
 with a finite effective
 value $K_a^*$. Therefore the correlation functions are again
 algebraic, containing this effective parameter.
 It is this second scenario that we are interested in, not the actual
 phase transition itself.
\begin{figure}
\includegraphics[width=4.8cm]{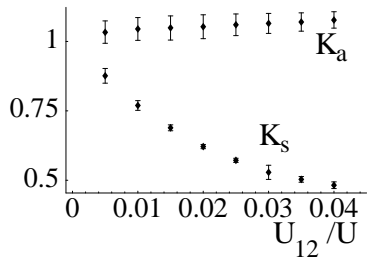}
\caption{\label{scale}
 LL parameters $K_s$ and $K_a$, as a function of $U_{12}$,
 for the same parameters as in Fig. \ref{density},
 from numerical fitting.
}
\end{figure}
%
%
%

 To determine the phase diagram 
 we consider the correlation functions of these
 order parameters:
 single-particle SF, described by $O_{SF}=b_a$,
 CDW order, corresponding to the $2 k_F$-component
 of the density operator $O_{CDW}=n_a$, 
 and paired SF, 
 $O_{PSF}=b_1 b_2$, which appears
 on the attractive side.
 The form of these correlation functions
 is 
  $\langle O(x)O(0)\rangle \sim |x|^{\alpha-2}$, except
 for the single-particle SF in the paired regime, where it decays
 exponentially.
  An order parameter $O(x)$ has QLRO, if its correlation function
 is algebraic, and $\alpha>0$. This implies that the corresponding
 susceptibility is divergent, indicating an instability towards ordering
 \cite{giamarchi_book}.
 The scaling exponent of $O_{SF}$ is 
%
$\alpha_{SF}  = 2- 1/4K_s - 1/4K_a$, 
 the one of $O_{CDW}$ is 
 $\alpha_{CDW} = 2-  K_s - K_a$, and PSF has
 $\alpha_{PSF}  = 2- 1/K_s$.
 We also consider the anti-pair operator $b_1^\dagger b_2$, 
 which has a scaling exponent of $2-1/K_a$.
 We use the latter two correlation functions in the numerical
 fitting procedure. 
%
%
%
%
 In Fig. \ref{PD} we see the
 resulting
 phase diagram.
 For attractive interactions we see the formation of a paired phase,
   in which two regimes of quasi-order are found:
 in the entire paired regime, PSF  
 is the dominant QLRO, whereas for part of that regime we find CDW
 as a subdominant order. The latter can be considered a 
 SS of
 pairs, whereas single-particle SF is destroyed.
 On the repulsive side we find the 
 SS phase that we look for in this
 paper.
  Using the 
  flow invariant $g_\sigma^2 - 4 \pi^2 (K_a -1)^2$, and 
  Eq. 
 (\ref{K}), we
 determine the nearly linear 
  SS phase boundary for small repulsive
 interactions to be:
 $U_{12}/v \geq 32 t \sin \pi n /U$. 

%

%
%
%
%
%
%
%

%
%
%
%
%
%
%

Having given the phase diagram of the infinite, homogeneous
  system,
 we now address the question of how supersolidity 
  can be found in actual cold atom systems.
 For this, we use a 
 TEBD method 
 \cite{vidal} to obtain 
 the ground state of the system. 
 %
 We 
  choose a small value for
 $t/U$, 
 to reach the near-hardcore regime,
 and a positive $U_{12}$ of the order of $t$, 
 to 
 avoid phase separation.
 We choose the atom number and the 
 global trap parameter, 
 such that
 the density is smaller than $0.5$ throughout the system.
  In Fig. \ref{density} we show the densities of the two species for 
 the case $t/U=0.005$, $U_{12}/U=0.04$, 
 $\Omega=10^{-5}U$, and a particle number of $17$ of
 each atom species, on a lattice of $90$ sites, as an illustration
 of the ground state. One can clearly see the density
 modulation of each species, whose
 wavelength is determined by the density, not the lattice.

A central question when studying `phases' in finite-size, non-homogeneous
 systems, is whether a given state can be reasonably related to a
  phase of the associated system in the thermodynamic limit.
 We address this question 
  by 
 numerically fitting the correlation functions of the
 pair and the anti-pair operator in the bulk of the system 
  with power-law functions.
 We find 
 a very good fit,
 and
 we depict the scaling exponents extracted from the numerical data in
 Fig. \ref{scale}, as a function of the interaction $U_{12}$.

 Having established that the state of the system is indeed a
  supersolid, we now turn to the crucial question
 of how this phase can be detected in experiment.
 We propose two measurements that would address this question:
 1) a TOF interference measurement 
  to determine the single-particle
 correlation function, which is the defining quantity of SF QLRO, and
 2) a measurement of the structure factor \cite{struct} to determine the
 density correlation function, which is the defining quantity of
 CDW QLRO.

%
%

%
%
\begin{figure}
\includegraphics[width=5.3cm]{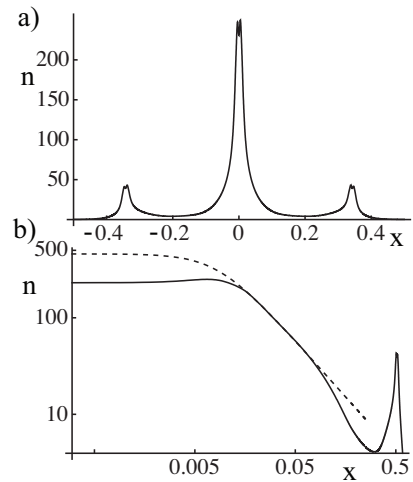}
\caption{\label{TOF}
Interference pattern of the 
 atomic 
 mixture, with the
 parameters of Fig. \ref{density}, when interpreted as $^{87}$Rb,
 released from an optical lattice with lattice constant $a=400$nm, localized
 states with a length scale $d=60$nm, after an expansion time $t = 30$ms.
 This could be realized with an optical lattice potential of $18 E_R$ in 
 longitudinal,
 and $30 E_R$ in transversal direction,
 $E_R$ being the recoil energy. 
 In a) we see the 1D density $n$ of the full pattern 
 (in units mm$^{-1}$), as a function of the spatial coordinate $x$ (in mm).
 In b) we fit the central interference peak with a power-law. The
 good numerical fit indicates the presence of single-particle
 quasi-superfluidity.
}
\end{figure}
%
%
%

{\it TOF measurement.}
  We assume that when the optical lattice is turned off, the
 atoms expand freely, that is 
%
$b_a(x,t)  =  \sum_j w(x-r_j)  b_{a,j}$, 
%
 where 
%
 $w(x,t)  = 
 \sqrt{d/\sqrt{2\pi} \Delta(t)^2}
 \exp(-x^2/4 \Delta(t)^2)$, 
%
%
%
  with $\Delta(t)^2 = d^2 + i t \hbar/2m$, 
 $a$ the lattice constant, $d$ the 
 width of the initial
 state, assumed Gaussian, $t$  the expansion time,
 and $m$ the atomic mass.
  We calculate the density 
 $n(x) = \langle b^\dagger(x,t) b(x,t) \rangle$,
 which contains the single particle correlation function of
 the original system,
 which we show in Fig. \ref{TOF}. 
  Fig. \ref{TOF} a) shows 
 the full interference pattern, and \ref{TOF} b)
  the central peak with a fit with a power-law
 function $c(x^2 + a^2)^{\alpha/2-1}$,
 shown on a log-log scale.
  The good agreement with power-law scaling
 indicates the presence of SF QLRO. 

%
%
\begin{figure}
\includegraphics[width=8.8cm]{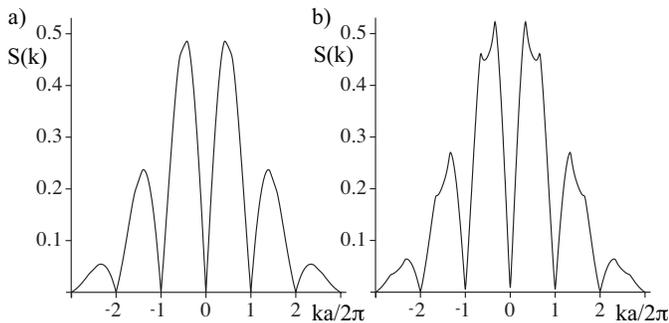}
\caption{\label{struct}
 Structure factor of the atomic mixture, with the same parameters
 as in Fig. \ref{TOF}, in 1/${\textstyle \mu}$m, as a function
 of $k$ in units of $2\pi/a$.
  a) corresponds to very weak interactions $U_{12}/U=0.0025$,
 and shows no particular structure beyond a SF signature,
 b) corresponds to $U_{12}/U=0.04$. We
 see additional peaks, that indicate CDW QLRO at a momentum 
 that is consistent with the density at the center of the trap. 
}
\end{figure}

{\it Structure factor.} 
 We assume, as an example, an in-situ measurement of the same system
 described before.
 The real-space structure factor is related to the density-density 
 correlation function of the lattice system by:
\bea
S(k) & \approx & \int \frac{d x_1 d x_2}{L} e^{-i k x_{12}}
(\langle n_{x_1} n_{x_2}\rangle -  \langle n_{x_1}\rangle
 \langle n_{x_2}\rangle)
\eea
with $x_{12}=x_1-x_2$, and the real-space correlation function defined as 
\bea
\langle n_{x_1} n_{x_2}\rangle & = & \sum_{i_1, i_2} |w(x_1 - r_{i_1})|^2  
 |w(x_2 - r_{i_2})|^2  
 \langle n_{i_1} n_{i_2}\rangle.\nonumber
\eea
%
%
 $\langle n_{x_1}\rangle \langle n_{x_2}\rangle$ is similarly defined.
 Fig. \ref{struct} shows $S(k)$ 
 for $U_{12}/U=0.04$, and
 a nearly non-interacting example,
 $U_{12}/U=0.0025$.
  The 
 envelope of the function is given by
 the inverse 
 width of the Wannier state;
 the periodic shape comes about because we map a lattice 
 quantity onto real space. 
 We clearly see the onset of a peaked structure at momenta that are
 consistent with the density at the trap center.
 These
 peaks are related to algebraic cusps in the static structure
 factor, when the CDW regime is reached, 
%
 $S(2 k_F + q)  \sim  |q|^{\alpha_{CDW}-1}$.
%
%
 A 
 measurement of the dynamic structure factor would lead
 to an even more striking result, since these cusps would translate
 into algebraic divergencies.
We also note that a measurement of the lattice density 
 in-situ\cite{nelson} could
 depict a profile as in Fig. \ref{density}, and that repeated
 measurements and a correlation analysis as in \cite{noise}, but
 without expansion, could give the real-space density correlation function 
 directly.

In conclusion, we have presented a proposal of how a supersolid 
 phase of binary bosonic mixtures in 1D can be created and probed in present ultracold atom experiments: in particular, with only local interatomic interactions.
 Using LL theory, we identified the generic phase diagram
 of this system for incommensurate filling; with
 TEBD simulations we found a concrete example of a finite, realistic system, including
 a global trap, which shows both SF and CDW QLRO.
 Two well-established
 measurement techniques, the TOF signal and the structure factor,
 provide clear experimental signatures of the two orders present in this remarkable state of matter. 
 
We appreciate conversations with Ian Spielman. 
 I. D. acknowledges support from a Grant-in-Aid from JSPS.

%

\def\etal{\textit{et al.}}


\begin{thebibliography}{99}

\bibitem{earlysupersolid} A. F. Andreev and I. M. Lifshitz, {\it Sov. Phys. JETP} {\bf 29}, 1107 (1969);
 C. V. Chester, {\it Phys. Rev. A} {\bf 2}, 256 (1970);
 A. J. Leggett, {\it Phys. Rev. Lett.} {\bf 25}, 1543 (1970).


\bibitem{supersolid}
 G.G. Batrouni et al., {\it Phys. Rev. Lett.} {\bf 97}, 087209 (2006);
 V.W. Scarola {\it et al.},
 {\it ibid.}
 {\bf 95},
033003 (2005);
P. Sengupta {\it et al.}, 
 {\it ibid.} {\bf 94}, 207202 (2005);
S. Wessel {\it et al.},
 {\it ibid.} 
 {\bf 95}, 127205 (2005); D. Heidarian {\it et al.},
 {\it ibid.}  {\bf 95}, 127206 (2005); R.G. Melko et al., 
 {\it ibid.} {\bf 95}, 127207 (2005);
 H.P. B\"{u}chler {\it et al.},
  {\it ibid.} {\bf 91}, 130404 (2004);
  M. Boninsegni {\it et al.},
  {\it ibid.} {\bf 95}, 237204 (2005);
  M. Boninsegni,
  J. Low Temp. Phys. {\bf 132}, 39 (2005);
D.L. Kovrizhin {\it et al.}, {\it Europhys. Lett.} {\bf 72}, 162 (2005);
 F. Karim Pour, {\it et al.}, Phys. Rev. B {\bf 75}, 161104(R) (2007).


\bibitem{commix}
L. Mathey,  {\it Phys. Rev. B} {\bf 75}, 144510 (2007).

\bibitem{kimchan} E. Kim and M. Chan, {\it Nature} {\bf 427}, 225 (2004).

\bibitem{kimchantheory}
A. Leggett, {\it Science} {\bf 305}, 1921 (2004); N. Prokof'ev and
B. Svistunov, {\it Phys. Rev. Lett.} {\bf 94}, 155302 (2005).


\bibitem{greiner}
M. Greiner, {\it et al.}, {\it Nature} {\bf 415}, 39 (2002).

\bibitem{BECBCS}
M. Greiner, {\it et al.}, {\it Nature} {\bf 426}, 537 (2003); S. Jochim, {\it et al.}, {\it Science} {\bf 302}, 2101 (2003); M.W. Zwierlein, {\it et al.},
 {\it Phys. Rev. Lett.} {\bf 91}, 250401 (2003).

\bibitem{noise}
M. Greiner, {\it et al.}, {\it Phys. Rev. Lett.} {\bf 94}, 110401 (2005);
S. Foelling, {\it et al.} {\it Nature} {\bf 434}, 481 (2005);
 E. Altman, {\it et al.}, {\it Phys. Rev. A} {\bf 70}, 013603 (2004);
 L. Mathey, {\it et al.}, cond-mat/0507108.

\bibitem{zoran}
Z. Hadzibabic, {\it et al.}, {\it Nature} {\bf 441}, 1118 (2006).

\bibitem{Tonks}
B. Paredes, {\it et al.}, {\it Nature} {\bf 429}, 277 (2004).

\bibitem{kinoshita}
T. Kinoshita, {\it et al.},  
 {\it Science} {\bf 305},
 1125 (2004).

\bibitem{1Dgas}
B. Laburthe Tolra, {\it et al.}, {\it Phys. Rev. Lett.} {\bf 92}, 190401 (2004);
 C. D. Fertig, {\it et al.}, {\it ibid.} {\bf 94}, 120403 (2005).


\bibitem{stoeferle}
T. St\"{o}ferle, {\it et al.}, {\it Phys. Rev. Lett.} {\bf 92}, 130403 (2004).

\bibitem{Ian}
I. B. Spielman, {\it et al.}, {\it Phys. Rev. Lett.} {\bf 98}, 080404 (2007);
 {\it ibid.} {\bf 100}, 120402 (2008).


\bibitem{vidal}
G. Vidal, {\it Phys. Rev. Lett.} {\bf 93}, 040502 (2004); 
  {\it ibid.} {\bf 98}, 070201 (2007).

\bibitem{Schulz93}
H. J. Schulz, {\it Phys. Rev. Lett.} {\bf 71}, 1864 (1993). 

\bibitem{Haldane}
F. D. M. Haldane, {\it Phys. Rev. Lett.} {\bf 47}, 1840 (1981).

\bibitem{Cazalilla}
M.A. Cazalilla, {\it J. Phys. B: At. Mol. Opt. Phys.} {\bf 37}, S1 (2004).
 
\bibitem{CazalillaTonks}
M.A. Cazalilla, {\it Phys. Rev. A} {\bf 70}, 041604(R) (2004).

\bibitem{giamarchi_book}
T. Giamarchi, {\it Quantum Physics in one dimension},(Oxford Univ. Press, Oxford, UK, 2004).

\bibitem{struct}
J. Stenger, {\it et al.}, {\it Phys. Rev. Lett.} {\bf 82}, 4569 (1999);
 J. Steinhauer,  {\it et al.}, {\it ibid.} {\bf 88}, 120407 (2002).

\bibitem{nelson}
K. D. Nelson, {\it et al.}, 
 {\it Nature Physics} {\bf 3}, 556 (2007).























\end{thebibliography}
\end{document}